\documentstyle[twocolumn,pra,aps]{revtex}

\begin{document}

\bibliographystyle{prsty}

\draft

%\preprint{}\omega = -\omega_0

\tighten

\title{Polarization fluctuations in vertical cavity surface emitting lasers:\\ a key to the mechanism behind polarization stability}

\author{Holger F. Hofmann and Ortwin Hess}
\address{Institut f\"ur Technische Physik, DLR\\
Pfaffenwaldring 38--40, D--70569 Stuttgart, Germany}

\date{\today}

\maketitle

\begin{abstract}
We investigate the effects of the electron-hole spin dynamics on the 
polarization fluctuations in the light emitted from a vertical cavity surface 
emitting laser (VCSEL). The Langevin equations are derived based on a rate equation model including birefringence, dichroism and two carrier density pools
separately coupled to right and left circular polarization. 
The results show that the carrier dynamics phase lock the polarization 
fluctuations to the laser mode. This is clearly seen in the difference 
between the fluctuations in ellipticity and the fluctuations in polarization 
direction. Separate measurements of the polarization fluctuations in 
ellipticity and in polarization direction can therefore provide quantitative
information on the non-linear contribution of the carrier dynamics to 
polarization stability in VCSELs.
\end{abstract}

%\pacs{PACS numbers:
%42.55.Px,  
%Semiconductor lasers; laser diodes.
%78.20.Bh  
%Theory, models, and numerical simulation
%}

\section{Introduction}

\label{sec:intro}

Although vertical cavity surface emitting lasers (VCSELs) are often highly
symmetric around the axis of laser emission, practical devices usually
emit linearily polarized light. The investigation of the weak anisotropies
responsible for this polarization stability has been the object of a number
of recent experimental and theoretical studies 
\cite{Cho94,Cho95,Jan96,Jan96a,Tra96,Mar96}. 
One of the questions raised in this search for a better understanding of the
polarization properties of VCSELs is whether it is sufficient to consider the
effects of the optical anisotropies separately from the carrier dynamics or
whether the highly anisotropic saturation of the gain in quantum well VCSELs
must be taken into account as well. This question is particularily
complicated because semiconductor lasers are type B lasers and 
usually it is not realistic to adiabatically eliminate the carrier dynamics
from the laser equations. The correct description of gain saturation effects 
therefore requires the inclusion of carrier densities as dynamical variables.
A rate equation model for the polarization dynamics of quantum well VCSELs  
including the spin dynamics of the carriers was introduced in 1995 by San 
Miguel and coworkers \cite{Mig95}. This model shows that the effects of the
linear anisotropies can be greatly modified as a result of the carrier 
dynamics \cite{Mar96}. In the case of relaxation oscillation frequencies 
greater than the spin relaxation rate the carrier dynamics can be observed
as relaxation oscillations in the fluctuations of ellipticity and of 
polarization direction \cite{Hof97}. Recent experimental results, however, 
suggest that the 
spin relaxation rate is likely to be greater than the relaxation oscillation 
frequency \cite{Jan96,Jan96a,Jan97}. 
In one study, the contributions of
the carrier dynamics have been neglected altogether since the observed 
polarization stability could be
interpreted in terms of a type A laser model including only linear optical
anisotropies \cite{Jan97}. A direct experimental determination of the carrier
dynamics contribution to polarization stability would thus be useful to
test the validity of the different models. 

In this paper we therefore investigate the different contributions of the
linear anisotropies to the polarization fluctuations in the case of fast 
spin relaxation. In section \ref{sec:model} the rate equations are introduced
and the Langevin equations at the stable point are formulated. In section
\ref{sec:fluct} the Langevin equation is solved and the resulting polarization
fluctuations are presented. In section \ref{sec:disc} the difference between the 
contributions of birefringence and dichroism and the contributions of the 
relaxation oscillation dynamics of the carriers is discussed and experimental 
possibilities of identifying the contributions are proposed.
In section \ref{sec:concl} the conclusions are presented. 

%----------------------------------------------------------------------

\section{Polarization dynamics of the split density model}

\label{sec:model}

\subsection{The rate equations}

In the model introduced by San Miguel \cite{Mig95}, the carrier density 
is subdivided into two carrier density pools interacting only with right or
left circular polarized light, respectively. The physical justification for
this assumption is the conservation of angular momentum around the axis of
symmetry. In the following equations, we will use the parameters $D$ for the 
total carrier density above transparency and $d$ for the difference between
the carrier density interacting with right circular polarized light and the
carrier density interacting with left circular polarized light. $n$ is the 
total number of photons in the cavity. The polarization is described using 
the normalized Stokes parameters. In terms of the complex amplitudes of the
circular polarized light field modes $E_+$ and $E_-$, these are
\begin{mathletters}
\begin{eqnarray}
\label{eq:stokespara}
P_1 & = & \frac{E_+^*E_- + E_-^*E_+}{E_+^*E_+ + E_-^*E_-}\\
P_2 & = & -i\frac{E_+^*E_- - E_-^*E_+}{E_+^*E_+ + E_-^*E_-}\\
P_3 & = & \frac{E_+^*E_+ - E_-^*E_-}{E_+^*E_+ + E_-^*E_-}.
\end{eqnarray}
\end{mathletters}
The relevant timescales of the laser process are given by the rate of 
spontaneous 
emission into the laser mode, $2w$  (usually around $10^6 - 10^7 s^{-1}$),
the rate of emission into non-laser modes $\gamma$ (usually around 
$10^9-10^{10} s^{-1}$), and the rate of emission from the cavity, $2\kappa$
(usually around $10^{12} - 10^{13} s^{-1}$). In addition, the spin flip 
scattering rate $\gamma_s$ is an important timescale for the polarization 
sensitive interaction between the light field and the carrier densities.
It is expected to be around $10^{10} - 10^{12} s^{-1}$. In light of the 
experimental results presented in \cite{Jan96,Jan96a,Jan97} we will
in the following consider $\gamma_s$ to be much faster than the relaxation 
oscillation and adiabatically eliminate the spin dynamics.
Note that the opposite case, i.e. where  $\gamma_s$ is much smaller
than the relaxation oscillation frequency, has 
been described in \cite{Hof97}.

Similar to the definition of the Stokes vector, the anisotropies can be
defined as vectors. The orientation of the vector indicates the polarization
for which the respective physical quantities are at a maximum. 
There are three types of anisotropies:
\begin{enumerate}
\item the relative gain anisotropy given
by {\bf g}, such that the rate of spontaneous emission into the laser mode
is $2w(1+{\bf P}\cdot {\bf g})$
\item the loss anisotropy {\bf l}, such that the 
rate of 
photon emission from the cavity is given by $2\kappa(1+{\bf P}\cdot{\bf l})$ 
\item
the frequency anisotropy ${\bf \Omega}$, such that the length of $\Omega$ is
equal to the frequency difference between the modes of orthogonal 
polarization. 
\end{enumerate}
Since we only consider small anisotropies, we will neglect 
the effects of gain and loss anisotropies on the total intensity of the
laser process by assuming that
$1+{\bf g}\cdot{\bf P} \approx 1$ and $1+{\bf g}\cdot{\bf P} \approx 1$. 
The rate equations are then given by   
\begin{mathletters}
\begin{eqnarray}
\label{eq:rateeqns}
\frac{d}{dt}D &=& -wDn - \gamma D              - wdnP_3 + \mu\\
\frac{d}{dt}n &=&  wDn -2\kappa n              + wdnP_3      \\
\frac{d}{dt}d &=& -wdn- (\gamma + \gamma_s) d  - wDnP_3      \\
\frac{d}{dt}{\bf P} &=& [(w(D{\bf g} + d\hat{{\bf e_3}}) 
                        - 2\kappa{\bf l})\times {\bf P}]\times{\bf P} 
\nonumber \\ && +({\bf \Omega} +w\alpha d \hat{{\bf e_3}})\times {\bf P} 
\end{eqnarray}
\end{mathletters}
$\hat{{\bf e_3}}$ indicates the unit vector in the direction of the 3rd component
of the stokes vector.

$\mu$ is the injection current above transparency and $\alpha$ is the 
linewidth enhancement factor which describes a shift in frequency due 
to the electron-hole density in the quantum well.
 
\subsection{Adiabatic elimination of the carrier density difference $d$}

If we assume that the spin flip scattering rate $\gamma_s$ is much faster 
than all the other timescales
involved in the laser process, we can adiabatically eliminate the carrier 
density difference $d$ by using

\begin{equation}
d= \frac{wDnP_3}{\gamma_s}
\end{equation}

We can also assume that $d$ will always be much smaller than the total
carrier density $D$, such that $wDn+wdnP_3\approx wDn$. The rate equations 
of the total carrier density $D$ and the total photon number $n$ are then
independent of the polarization dynamics.

\begin{mathletters}
\begin{eqnarray}
\label{eq:adiabat}
\frac{d}{dt}D &=& -wDn - \gamma D + \mu\\
\frac{d}{dt}n &=&  wDn -2\kappa n      \\
\frac{d}{dt}{\bf P} &=& [(wD({\bf g} + \frac{wnP_3}{\gamma_s} \hat{{\bf e_3}}) 
                        - 2\kappa{\bf l})\times {\bf P}]\times{\bf P} 
\nonumber \\ &&
+({\bf \Omega} +wD\alpha \frac{wnP_3}{\gamma_s} \hat{{\bf e_3}})\times {\bf P} 
\end{eqnarray}
\end{mathletters}
In order to describe the polarization dynamics at constant laser intensity 
the stationary solutions of the total carrier density $D = 2\kappa/w$ and the 
stationary photon number $n$ may be applied to the polarization dynamics. The 
equation for the Stokes parameter dynamics then reads
\begin{eqnarray}
\label{eq:optic}
\frac{d}{dt}{\bf P} &=& [({\bf s} + \frac{2\kappa wn}{\gamma_s} P_3
                        \hat{{\bf e_3}})\times {\bf P}]\times{\bf P} 
\nonumber \\ &&         +({\bf \Omega}+\alpha\frac{2\kappa wn}{\gamma_s}
                        P_3\hat{{\bf e_3}})\times {\bf P} 
\end{eqnarray}
with ${\bf s} = 2\kappa ({\bf g}-{\bf l})$ as the total dichroism. Note that 
for $2\kappa wn/\gamma_s = 0$ this equation is essentially the Stokes 
parameter version of the linear equation used in \cite{Jan97}. The
interpretation of the polarization stability in terms of a type A laser model
is possible because $d$ can be adiabatically eliminated. It should be noted, 
however, that the type A model does not correctly describe the dynamics of 
the field intensity. The use of normalized Stokes parameters is therefore
more appropriate. In particular, the possibility of relaxation oscillations in
the laser intensity is not included in the equations used in \cite{Jan97}.

The effects of the split density model are given by the terms proportional to
$2\kappa wn/\gamma_s$. 
$2\kappa wn$ is approximately equal to the square of the 
relaxation oscillation frequency. To estimate the importance of the split 
density model contributions to the laser dynamics it is therefore useful to
compare the relaxation oscillation frequency with the spin flip scattering 
rate.  

\subsection{Langevin equation}

In the case of stable linear polarization with parallel birefringence 
${\bf \Omega} = \Omega \hat{{\bf e_1}}$ and dichroism  ${\bf s} = 
s \hat{{\bf e_1}}$ the stationary Stokes vector is ${\bf P} = \hat{{\bf e_1}}$.
Fluctuations of ${\bf P}$ are given by the components $P_2$ for fluctuations
in polarization direction and $P_3$ for fluctuations in ellipticity.
The linearized Langevin equation for $P_2$ and $P_3$ derived from equation
(\ref{eq:optic}) is 
\begin{equation}
\label{eq:langevin}
\frac{d}{dt}\left(
\begin{array}{c}
P_2\\P_3
\end{array}\right) = \left(\! 
\begin{array}{c@{\hspace{0.5 cm}}c}
-s & -\Omega-\alpha\chi n \\
\Omega & -s -\chi n
\end{array}\right)
\left(
\begin{array}{c}
P_2\\P_3
\end{array}\right) + {\bf f}(t)
\end{equation}
where $\chi = 2\kappa w/\gamma_s$ is the coefficient of the laser intensity
dependent contribution of the carrier dynamics in the split density model.

The contributions of dichroism $s$, birefringence $\Omega$ and carrier 
dynamics $\chi n$ can be
recognized clearly in this matrix equation. The noise term ${\bf f}(t)$ is
a consequence of the
vacuum fluctuations in the electromagnetic field entering the cavity and in 
the dipole density of the gain medium. 
The magnitude of the noise terms for the Langevin equations may be derived by 
considering that photonic shot noise must be present both in the 
circular polarized modes and in the linear polarized modes at 45 degrees
to the polarization of the laser light. 

\begin{mathletters}
\begin{eqnarray}
\label{eq:influc1}
\langle f_{p_2}(t)f_{p_2}(t+\tau)\rangle 
&=& \frac{4\kappa}{n}\delta(\tau)\\
\label{eq:influc2}
\langle f_{p_3}(t)f_{p_3}(t+\tau)\rangle 
&=& \frac{4\kappa}{n}\delta(\tau).
\end{eqnarray}
\end{mathletters}
The factor of $4\kappa/n$ is the rate $4\kappa n$ at which photons enter
and leave the 
cavity, divided by the squared normalization of the Stokes parameter.
This is the minimum noise term necessary to satisfy the quantum mechanical
uncertainty relations. Additional noise may arise from reabsorption of
photons into the laser medium due to incomplete inversion. This effect will 
be extremely strong in ultra low threshold lasers \cite{Bjo94}. For typical 
VCSELs however, the minimal noise 
terms should be the main contribution to ${\bf f}(t)$.  

%----------------------------------------------------------------------

\section{Solution of the Langevin equation}

\label{sec:fluct}

\subsection{Linear response near the stationary point}

The eigenvalues $\lambda_\pm$ and the left and right eigenvectors 
${\bf a_{\pm}}$ and ${\bf b_{\pm}}$ of the
$2 \times 2$ matrix describing the relaxation dynamics of polarization
fluctuations in equation (\ref{eq:langevin}) can be analytically determined
to obtain the linear response of the laser to the polarization fluctuations.  

\begin{mathletters}
\begin{equation}
\lambda_{\pm} = -s -\frac{\chi n}{2} 
                \pm i \Omega \sqrt{1 + \frac{\alpha\chi n}{\Omega}
                            -(\frac{\chi n}{2\Omega})^2}  
\end{equation}
\begin{equation}
{\bf a}_{\pm}=\frac{1}{\sqrt{2}}
\left(
\begin{array}{c@{\hspace{0.5cm}}c}
1, & \frac{\chi n}{2 \Omega} 
    \pm i \sqrt{1+\frac{\alpha\chi n}{\Omega}-(\frac{\chi n}{2\Omega})^2}
\end{array}\right)
\end{equation}
\begin{equation}
{\bf b}_{\pm}=\frac{1}{\sqrt{2}}
\left(
\begin{array}{c}
1 \\ \\ 
\frac{\Omega}{\Omega + \alpha\chi n} \left(\frac{\chi n}{2 \Omega} 
\mp i \sqrt{1+\frac{\alpha\chi n}{\Omega}-(\frac{\chi n}{2\Omega})^2}\right) 
\end{array}\right)
\end{equation}
\end{mathletters}
The eigenvalues already indicate an intensity dependent contribution to both 
the frequency and the relaxation rate of polarization fluctuations. 
The linear response to the polarization fluctuations is then given by the
Green´s function 

\begin{equation}
{\bf G}(\tau) = e^{\lambda_+\tau}{\bf b}_+\otimes{\bf a}_+
               +e^{\lambda_-\tau}{\bf b}_-\otimes{\bf a}_-.
\end{equation}
The fluctuations in the polarization of the laser light can be determined
by applying this Green´s function to the noise term ${\bf f}(t)$ in the
Langevin equation.

\subsection{Polarization fluctuations}

The fluctuations of polarization direction $P_2$ and ellipticity $P_3$,
as well as their correlations are given by the following correlation 
functions:
\begin{mathletters}
\begin{eqnarray}
&&\langle P_2(t)P_2(t+\tau)\rangle =
\frac{4\kappa}{(2s+\chi n)n}
\nonumber \\ && \hspace{1cm}
(1+\frac{\alpha\chi n}{2\Omega}) e^{-(s+\chi n/2)\tau}
\cos(\omega_0\tau)\\ &&
\langle P_3(t)P_3(t+\tau)\rangle =
\frac{4\kappa}{(2s+\chi n)n}
\nonumber \\ && \hspace{1cm}
(1+\frac{\alpha\chi n}{2\Omega}) e^{-(s+\chi n/2)\tau}
\frac{\Omega}{\Omega+\alpha\chi n}\cos(\omega_0\tau)\\ &&
\langle P_2(t)P_3(t+\tau)\rangle =
\frac{4\kappa}{(2s+\chi n)n}(1+\frac{\alpha\chi n}{2\Omega}) e^{-(s+\chi n/2)\tau}
\nonumber \\ && \hspace{1cm}
(\frac{\chi n/2}{\Omega+\alpha\chi n}\cos(\omega_0\tau)
+ \frac{\omega_0}{\Omega+\alpha\chi n}\sin(\omega_0\tau))\\ &&
\langle P_3(t)P_2(t+\tau)\rangle =
\frac{4\kappa}{(2s+\chi n)n}(1+\frac{\alpha\chi n}{2\Omega}) e^{-(s+\chi n/2)\tau}
\nonumber \\ && \hspace{1cm}
(\frac{\chi n/2}{\Omega+\alpha\chi n}\cos(\omega_0\tau)
- \frac{\omega_0}{\Omega+\alpha\chi n}\sin(\omega_0\tau)).
\end{eqnarray}
\end{mathletters}
The frequency $\omega_0$ is given by the imaginary part of the eigenvalues,
\begin{equation}
\omega_0 = \Omega\sqrt{1 + \frac{\alpha\chi n}{\Omega}
                      -(\frac{\chi n}{2\Omega})^2}.
\end{equation}
Figure \ref{fluct} shows the two-time correlations as a function of the
delay time $\tau$ for a typical choice of parameters. 

In addition to the quantitative changes in the relaxation rate and the 
oscillation 
frequency, the non-linear contribution of the carrier dynamics introduce
clear qualitative modifications to the polarization fluctuations. 
This is a consequence of the split density model in which the ellipticity
is stabilized by the carrier dynamics in addition to the dichroism. 
Therefore the fluctuations in ellipticity are smaller than the fluctuations
in polarization direction. Also, the phase shift between the oscillations
of ellipticity and of polarization direction is not $\pi/2$ as one would expect
for a purely linear birefringence. These differences between a linear optical
model and the split density model may provide possibilities for experimental
investigations of the split carrier density contributions to polarization
stability. 

\subsection{Polarization noise in the emission spectrum}

Because the laser field amplitude $E_\parallel$ is large compared to the noise,
the two-time correlation function of the field amplitude polarized orthogonally
to the laser mode $E_\perp$ may be determined from the two-time correlations 
of the 
normalized Stokes parameters. In fact, the measurement of the Stokes parameters
corresponds to a heterodyne detection of the field dynamics of $E_\perp$.
Consequently, the two-time correlations of $E_\perp$ may be obtained using

\begin{equation}
E_\perp = \frac{P_2+iP_3}{2}E_\parallel .
\end{equation}

The measurement of the fluctuations in ellipticity and in polarization
direction is therefore equivalent to a phase sensitive measurement of the
fluctuations in the laser cavity mode of orthogonal polarization to the 
lasing mode. The difference between the in-phase fluctuations corresponding
to the polarization direction and the out-of-phase fluctuations corresponding
to ellipticity corresponds to a phase locking effect between the emissions
into the non-lasing polarization mode and the lasing mode.

The two-time correlation function of $E_\perp$ also shows some features of
the phase locking. However, the information is somewhat hidden by the summation
over the separate contributions. Neglecting the fluctuations in $E_\parallel$,
the two-time correlation function of $E_\perp$ is
\begin{eqnarray}
\langle E_\perp^*(t)E_\perp(t+\tau)\rangle &=& 
   \frac{n}{4}(   \langle P_2(t)P_2(t+\tau)\rangle  \nonumber \\
              &+& \langle P_3(t)P_3(t+\tau)\rangle  \nonumber \\ 
            &+ i& \langle P_2(t)P_3(t+\tau)\rangle  \nonumber \\
            &- i& \langle P_3(t)P_2(t+\tau)\rangle) \nonumber 
\end{eqnarray}
\begin{eqnarray}
     =& \frac{2\kappa(\Omega+\alpha\chi n/2)}{(2s+\chi n)(\Omega+\alpha\chi n)}
         ((1+\frac{\alpha\chi n}{2\Omega})\cos (\omega_0\tau)&
\nonumber \\ & + i \frac{\omega_0}{\Omega}\sin (\omega_0\tau))
         e^{-(s+\chi n/2)\tau} &\nonumber \\[0.5cm]
     =& \frac{2\kappa(\Omega+\alpha\chi n/2)}{(2s+\chi n)(\Omega+\alpha\chi n)}
             (\frac{\omega_0 + \Omega + \alpha\chi n/2}{2\Omega}
               e^{+i\omega_0\tau}& \nonumber \\
             & - \frac{\omega_0 - \Omega - \alpha\chi n/2}{2\Omega}
               e^{-i\omega_0\tau}
               )e^{-(s+\chi n/2)\tau}.&\\[0.3cm]\nonumber
\end{eqnarray}
The Intensity spectrum 
$I_\perp(\omega)=\langle E_\perp^*(\omega)E_\perp(\omega)\rangle$
is the Fourier transform of the two-time correlation function. It is given by
two Lorentzians,

\begin{eqnarray}
\label{eq:spect}
I_\perp(\omega)
             &=\frac{2\kappa(\Omega+\alpha\chi n/2)}
                    {(2s+\chi n)(\Omega+\alpha\chi n)}& 
                  \nonumber \\
             &(\frac{\omega_0 + \Omega + \alpha\chi n/2}{2\Omega}&
               \frac{s+\chi n/2}{\pi((s+\chi n/2)^2+(\omega+\omega_0)^2)}
                  \nonumber \\
             &-\frac{\omega_0 - \Omega - \alpha\chi n/2}{2\Omega}&
               \frac{s+\chi n/2}{\pi((s+\chi n/2)^2+(\omega-\omega_0)^2)}
               ).\\[0.3cm]\nonumber
\end{eqnarray}
An example of such a spectrum is given in figure \ref{spect} using the same
parameters as for the two-time correlations shown in figure \ref{fluct}.
Note that the intensity is given in units of photon number inside the cavity. 
The intensity emitted is given by $2\kappa$ times this value. 
The minimum noise asumption of equations (\ref{eq:influc1}) 
and (\ref{eq:influc2}) may
be tested by comparing the total intensity in the orthogonally polarized mode
with the predictions of equation (\ref{eq:spect}).

For $\chi \neq 0$ the theory predicts not only the peak at $-\omega_0$ from
the laser line but also a much smaller peak at $+\omega_0$. To clarify the 
quantitative relation between the two peaks it is useful to treat 
$\chi n/\Omega$ as
a small perturbation. 
\begin{eqnarray}
&I_\perp(\omega)\approx& \nonumber \\
          &\frac{2\kappa}{2s+\chi n} (
           \frac{s+\chi n/2}{\pi((s+\chi n/2)^2+(\omega+\omega_0)^2)} &
           \nonumber \\
          & + (\frac{\chi n}{4\Omega})^2(\alpha^2+1)
             \frac{s+\chi n/2}{\pi((s+\chi n/2)^2+(\omega-\omega_0)^2)}
             )&.\\[0.3cm]\nonumber
\end{eqnarray}

If $\chi n/\Omega$ is large enough the small noise peak at the opposite side 
of the laser line should be sufficient for a determination of $\chi$. The fact
that no such peak was observed in \cite{Jan96,Jan96a,Jan97} indicates that
$\chi n$ is indeed small compared to $\Omega$ in the devices studied.

%----------------------------------------------------------------------

\section{Experimental possibilities}

\label{sec:disc}

\subsection{Spectrum of light polarized orthogonally to the lasing mode}

If the carrier density dynamics is negligible for polarization stability the
birefringence $\Omega$ is given by the frequency difference $\omega_0$ 
betweeen the
laser line and the emission line of the orthogonally polarized mode and the 
dichroism $s$ is given by one half of the linewidth at half maximum 
$\Delta\omega_{FWHM}$. This is the assumption used in \cite{Jan97}. That the 
linewidths reported in that paper seem to be larger than $2s$
may be a consequence of the carrier density dynamics. In particular, the 
linewidth with carrier density dynamics is given by
\begin{equation}
\Delta\omega_{FWHM}= 2s + \chi n.
\end{equation}
The frequency shift $\omega_0$ is much harder to identify. According to the 
theory the
frequency should increase with intensity until $\chi n = 2\alpha\Omega$ and
then decrease again until overdamping occurs at $\chi n = 2\Omega(\alpha +
\sqrt{\alpha^2+1})$. The linear increase of frequency is 
\begin{equation}
\omega_0 \approx \Omega + \alpha \chi n/2.
\end{equation}
This frequency shift has first been derived by van~der~Lem and Lenstra 
\cite{Lem97} who argue that the shift is an alpha-enhanced saturation effect.

The most important feature predicted by the split density model is the peak
on the opposite side of the laser line. The ratio of the peak intensity at
$\omega=\omega_0$ and at $\omega=-\omega_0$ is approximately given by 
\begin{equation}
\frac{I_\perp(\omega_0)}{I_\perp(-\omega_0)} \approx 
(\frac{\chi n}{4\Omega})^2(\alpha^2+1).
\end{equation}  
If the frequency shift $\omega_0$, the linewidth $\Delta\omega_{FWHM}$ and
the intensity ratio $I_\perp(\omega_0)/I_\perp(-\omega_0)$ are measured it
is possible to determine the contribution of the carrier density dynamics 
$\chi n$
as well as the dichroism $s$ and the birefringence $\Omega$ at a fixed laser
intensity. Note that to obtain correct quantitative results the linewidth 
enhancement factor $\alpha$ must be known.
The peak at $\omega = +\omega_0$ may be very small because of
the second order dependence on $\chi n/\Omega$. For $\chi n/\Omega = 0.1$ the
intensity in that peak would be 100 to 1000 times lower than that
in the main peak at $\omega = -\omega_0$. This may be the 
reason why it has not been observed in previous experiments. If there really 
is no peak at $\omega = +\omega_0$ however, it must be concluded that the split
density model is not a valid description of the polarization dynamics in the
device under consideration.

\subsection{Separate measurements of the fluctuations in ellipticity and 
in polarization direction}

While it is possible to observe the effects of phase locking between the 
$E_\perp$ mode and the lasing mode $E_\parallel$ in the spectrum by measuring
the small peak at the opposite side of the laser peak separate measurements of
the fluctuations in ellipticity $P_3$ and in polarization direction $P_2$ 
are more sensitive to the effects of $\chi n$ and reveal more details of the
phase locking. This is most clearly seen in figures \ref{fluct} and \ref{spect}.
At $\chi n/\Omega = 0.5$ the intensity at $\omega = +\omega_0$ is still very
small while the features of the polarization fluctuations clearly reveal
strong effects of the carrier dynamics. 

The most important indicator is the ratio of fluctuations
in polarization direction and in ellipticity. This ratio can even be measured
at low time resolutions since it is a constant over all times and frequencies.

\begin{equation}
\frac{\langle P_2(t)P_2(t+\tau)\rangle}{\langle P_3(t)P_3(t+\tau)\rangle}
= 1 + \frac{\alpha\chi n}{\Omega}.
\end{equation}

Note that the $\alpha$ factor is responsible for the different magnitude of
the fluctuations. Without the effects of the $\alpha$ factor the phase locking
effects would only appear in the correlations between ellipticity and 
polarization direction $\langle P_2(t)P_3(t+\tau)\rangle$ and 
$\langle P_3(t)P_2(t+\tau)\rangle$. If these correlations
are measured it is interesting to determine the phase shift $\delta \phi$ 
with

\begin{eqnarray}
\delta\phi &=& \arctan 
  (\frac{\langle P_2(t)P_3(t+\tau)\rangle + \langle P_3(t)P_2(t+\tau)\rangle}
        {\langle P_2(t)P_3(t+\tau)\rangle - \langle P_3(t)P_2(t+\tau)\rangle})
       \nonumber \\
           &=& \arctan
   (\frac{\chi n}{\sqrt{4\Omega^2 + 4\alpha\chi n \Omega -(\chi n)^2}})
       \nonumber \\
           &\approx& \frac{\chi n}{2\Omega}.
\end{eqnarray}

This phase shift shows how the additional damping of the ellipticity by the
carrier density dynamics makes the ellipticity fluctuate less than $\pi/2$
out of phase with the polarization direction. 
The frequency $\omega_0$, the relaxation rate $s+\chi n/2$,
the fluctuation ratio 
$\langle P_2(t)P_2(t+\tau)\rangle/\langle P_3(t)P_3(t+\tau)\rangle$ and the
correlation phase $\delta \phi$ provide all the information needed to 
determine the dichroism $s$, the birefringence $\Omega$, the contribution of 
the carrier density dynamics $\chi n$ and the linewidth enhancement factor 
$\alpha$.
Note that the time resolution necessary to measure the dynamics of polarization
fluctuations in this regime is given by the birefringence $\Omega$. This 
suggests that a time resolution of several nanoseconds may actually be 
sufficient.
 
%----------------------------------------------------------------------

\section{Conclusions}

\label{sec:concl}

The calculations presented here clearly demonstrate how the carrier dynamics
modify the polarization fluctuations both in the spectrum of $E_\perp$
and in the Stokes parameters of ellipticity $P_3$ and of polarization 
direction $P_2$. 
Even if the contribution of the carrier density dynamics $\chi n$ is very
small compared to the birefringence $\Omega$ a careful analysis of the 
experimental data on polarization fluctuations should reveal these 
modifications. Such an analysis would show whether 
an interpretation of polarization stability in terms of birefringence and
dichroism only is valid or not. If the carrier dynamics of the split density
model are relevant to the polarization stability of a given device, the
measurement of polarization fluctuations provides information on the
optical anisotropies $\Omega$ and $s$, on the spin relaxation rate $\gamma_s$
and on the linewidth enhancement factor $\alpha$. If the non-linear 
contribution $\chi$ turns out to be zero, it seems likely that valence bands
with an angular momentum other than $3/2$ around the axis perpendicular to
the quantum well also contribute to the laser process in vertical cavity
surface emitting lasers.

% \section*{Acknowledgments}

%=========================================================

%
%=========================================================

\begin{figure}
\caption{Two-time correlations of the polarization fluctuations  
$\langle P_i(t)P_j(t+\tau)\rangle$ 
for $\alpha =2$ , $s=0.5$ GHz, $\Omega=4$ GHz and $\chi n=2$ GHz. 
(a) shows the fluctuations in ellipticity $\langle P_3(t)P_3(t+\tau)\rangle$ 
and the fluctuations in polarization direction 
$\langle P_2(t)P_2(t+\tau)\rangle$. 
(b) shows the correlations between ellipticity
and polarization direction, $\langle P_2(t)P_3(t+\tau)\rangle$ and
$\langle P_3(t)P_2(t+\tau)\rangle$.}
\label{fluct}
\end{figure}

\begin{figure}
\caption{Spectrum $I_\perp(\omega)$ of the emission polarized orthogonally
to the lasing mode for $\alpha =2$ , $s=0.5$ GHz, $\Omega=4$ GHz and 
$\chi n=2$GHz. Even though $\chi n/\Omega = 0.5$ the peak near $+5$ GHz 
is still very small.}
\label{spect}
\end{figure}
  
%
%=============================================================

\end{document}